\documentclass[aps,prl,twocolumn,preprintnumbers,groupedaddress]{revtex4}

\usepackage{amssymb}
\usepackage{epsfig} 
\usepackage{amsmath,color} 
\usepackage{graphicx} 

\usepackage{wrapfig}

\newcommand{\beq}{\begin{equation}}
\newcommand{\eeq}{\end{equation}}
\newcommand{\bea}{\begin{eqnarray}}
\newcommand{\eea}{\end{eqnarray}}
\newcommand{\ba}{\begin{array}}
\newcommand{\ea}{\end{array}}
\newcommand{\bi}{\begin{itemize}}
\newcommand{\ei}{\end{itemize}}
\newcommand{\bn}{\begin{enumerate}}
\newcommand{\en}{\end{enumerate}}
\newcommand{\bc}{\begin{center}}
\newcommand{\ec}{\end{center}}
\renewcommand{\l}{\left}
\renewcommand{\r}{\right}

\newcommand{\eq}[1]{Eq.~(\ref{#1})}
\newcommand{\eqs}[2]{Eqs.~(\ref{#1}) and (\ref{#2})}

\newcommand{\GeV}{\mathinner{\mathrm{GeV}}}

\begin{document}

\preprint{FTUV-14-12-31}
\preprint{IFIC-14-86}

\title{Spiral Inflation with Coleman-Weinberg Potential}


\author{Gabriela Barenboim}
\email[]{Gabriela.Barenboim@uv.es}
\author{Wan-Il Park}
\email[]{Wanil.Park@uv.es}
\affiliation{
Departament de F\'isica Te\`orica and IFIC, Universitat de Val\`encia-CSIC, E-46100, Burjassot, Spain}


\date{\today}

\begin{abstract}
We apply the idea of spiral inflation to Coleman-Weinberg potential, and show that inflation matching well observations is allowed for a symmetry-breaking scale ranging from an intermediate scale to GUT scale even if the quartic coupling $\lambda$ is of $\mathcal{O}(0.1)$.
The tensor-to-scalar ratio can be of $\mathcal{O}(0.01)$ in case of GUT scale symmetry-breaking.
\end{abstract}


\maketitle


\section{Introduction}

In modern cosmology, the concept of inflation \cite{Guth:1980zm} is essential to understand the very early history of the universe.
Inflation solves various problems of the old Big-Bang cosmology, for example, the horizon- , flatness-, and unwanted relic-problems \cite{Guth:1980zm,Guth:1979bh}.
It also provides a very compelling seed of density perturbations in the present universe via the classicalized quantum fluctuations of the inflaton which is typically a scalar field \cite{Mukhanov:1981xt,Mukhanov:1982nu}.

In order to realize a sufficiently prolongued inflationary period giving a quasi scale invariant (but slightly red-tilted) power spectrum so as to match the recent precision observations \cite{Ade:2013zuv}, the inflaton-potential should be flat enough. 
Coleman-Weinberg (CW) potential \cite{Coleman:1973jx} is a good candidate for this purpose and has been considered at the early stage of inflation cosmology \cite{Albrecht:1982wi}.
The benefit of the CW potential as the inflaton-potential is that it can be directly connected to GUT scale or a low-energy symmetry breaking.
However, as recently studied in Ref.~\cite{Barenboim:2013wra}, it does not match well with the recent CMB observations by Planck satellite \cite{Ade:2013zuv}, and requires an extremely small quartic coupling of $\mathcal{O}(10^{-14})$ even for a Planckian symmetry breaking scale.

Generically, when considered in one dimensional field space, the requirement of a flat enough inflaton-potential faces severe theoretical difficulties. 
Notorious examples are the $\eta$-problem and the validity of effective field theory in regards of a trans-Planckian excursion.
In the last several years, interesting ideas have been developed to overcome such theoretical difficulties by extending the inflaton trajectory into at least two-dimensional field space so as to compactify the long trajectory into a sub-Planckian field space (see for example, \cite{Silverstein:2008sg,McAllister:2008hb,Berg:2009tg,McDonald:2014oza,McDonald:2014nqa,Li:2014vpa,Carone:2014cta}).
Also, in a similar line, very recently, we have proposed a simple and novel scenario of spiral inflation realized in a symmetry-breaking potential \cite{Barenboim:2014vea}.
The idea of spiral inflation can drastically change the conventional approach to inflation with CW potentials although its full UV realization is a non-trivial task.

In this paper, we apply the idea behind spiral inflation to the CW potential and show that inflation matching perfectly the Planck satellite observations can be realized even if the quartic coupling is of $\mathcal{O}(0.1)$ for a symmetry breaking scale ranging from an intermediate to GUT scales.
We also show that the tensor-to-scalar ratio can be of $\mathcal{O}(0.01)$ for GUT scale symmetry-breaking.

\section{The model}
Following our recent work \cite{Barenboim:2014vea}, we consider a complex field $\Phi \equiv \phi e^{i \theta}/\sqrt{2}$ having a potential \footnote{There can be a Hubble scale negative mass term associated with the $\eta$-problem, but it does not affect our argument.
So, we ignore it for simplicity.}:
\beq \label{V}
V = V_\phi + V_{\rm M}
\eeq
with 
\bea \label{VCW}
V_\phi &=& \lambda \phi^4 \l[ \ln \l( \phi/\phi_0 \r) - 1/4 \r] + \lambda \phi_0^4/4
\\ \label{VM}
V_{\rm M} &=& \Lambda^4 \l[ 1 - \sin(\phi/M + \theta) \r] 
\eea
where $\lambda$ is a dimensionless coupling, $\phi_0$ is the vacuum expectation value of $\phi$, $\Lambda$ and $M$ are mass scales that will be constrained by inflationary phenomenology, and $\theta = {\rm Arg}(\Phi)$.
We assume $V_0 \equiv \lambda \phi_0^4/4 \gg \Lambda^4$.

The potential in \eq{VM} may have a nonpertubative origin, as discussed in Refs. \cite{Silverstein:2008sg,McAllister:2008hb,Berg:2009tg,McDonald:2014nqa}, and gives a modulation to the simple CW potential in \eq{VCW}, resulting in a spiraling-out valley in the potential.
Depending on the specific values of $\Lambda$ and $M$, there is a possibility to achieve spiral inflation.
In such a case, when $\phi_0$ is close to Planck scale, we expect inflationary observables similar to Ref.~\cite{Barenboim:2014vea}.
So, here we consider the possibility of spiral inflation with $\phi_0$ similar to or well below the GUT scale.

\section{Inflation}
In the field basis, the equations of motions are given by
\bea \label{phi-eom}
0 &=& \ddot{\phi} + 3 H \dot{\phi} + \frac{\partial V}{\partial \phi}
\\ \label{theta-eom}
0 &=& \ddot{\theta} + \frac{\dot{\phi} \dot{\theta}}{\phi} + 3 H \dot{\theta} + \frac{\partial V}{\phi^2 \partial \theta}
\eea
Starting from around the top of the potential, inflaton is expected to trace closely the minimum of the valley where $\partial V/ \partial \phi = 0$ gives
\beq \label{dVmin}
\frac{\Lambda^4}{M} \cos \theta_\phi = 4 \lambda \phi^3 \ln \l( \frac{\phi}{\phi_0} \r)
\eeq
leading to 
\beq \label{dphi-dtheta}
d\phi 
= \frac{(\Lambda^4/M) \sin \theta_\phi}{4 \lambda \phi^2 \l[ 3 \ln \l(\frac{\phi_0}{\phi}\r) - 1 \r] - \l(\frac{\Lambda^4}{M^2}\r) \sin \theta_\phi} d\theta
\eeq
with $\theta_\phi \equiv \phi/M + \theta$.
Note that during inflation, for the cosmologically relevant scales, we expect $\phi \ll \phi_0$ and $\sin \theta_\phi \approx 1$ at the minimum of the valley.
Hence, \eq{dphi-dtheta} can be approximated as
\beq \label{dphi-dtheta-app}
d\phi \approx - M d \theta
\eeq
In the flat field space, the infinitisimal displacement of inflaton is related to the ones of $\phi$ and $\theta$ as  
\beq \label{dI}
dI^2 = \l( d\phi \r)^2 + \l( \phi d\theta \r)^2 
\eeq 
and one finds
\bea
\frac{\partial I}{\partial \phi} &=& \l[ \l(\frac{\phi d\theta}{d\phi}\r)^2 + 1 \r]^{1/2}
\\
\frac{\partial I}{\partial \theta} &=& - \l[ \l(\frac{d\phi}{\phi d\theta}\r)^2 + 1 \r]^{1/2} \phi
\eea
At the minimum, the elements of the mass matrix are given by
\bea
\frac{\partial^2 V}{\partial \phi^2} &=& \frac{\Lambda^4}{M^2} \sin \theta_\phi - 4 \lambda \phi^2 \l[ 3 \ln \l( \frac{\phi_0}{\phi} \r) - 1 \r]
\\
\frac{\partial^2 V}{\phi \partial \theta \partial \phi} &=& \frac{\Lambda^4}{\phi M} \sin \theta_\phi
\\
\frac{\partial^2 V}{\phi^2 \partial \theta^2} &=& \frac{\Lambda^4}{\phi^2} \sin \theta_\phi
\eea
Defining 
\beq \label{a-delta}
a \equiv \frac{\phi}{M}, \quad \delta \equiv \frac{4 \lambda \phi^3 M \l[ 3 \ln \l( \frac{\phi_0}{\phi} \r) - 1 \r]}{\Lambda^4 \sin \theta_\phi}
\eeq
one finds that, for the cosmologically relevant scales with $a\gg \delta \sim \mathcal{O}(1)$ that will be justified in subsequent arguments, the mass eigenvalues of the two orthogonal directions can be approximated as
\bea \label{mIsq}
m_\parallel^2 &\approx& - \frac{4 \lambda \phi^2 \l[3 \ln \l(\frac{\phi_0}{\phi} \r) - 1\r]}{a \l( a - \delta + 1/a \r)} 
\\
m_\perp^2 &\approx& \l( a - \delta + \frac{1}{a} \r) \frac{\Lambda^4}{\phi M} \sin \theta_\phi 
\eea 
The inflaton is expected to follow the tachyonic direction.
Note that $m_\parallel$ is suppressed relative to the case of pure CW potential by the factor of $(M/\phi)^2$.

Even though inflation takes place in a $2$-dimensional field space, it behaves as in the single field case.
Hence, in terms of slow-roll parameters, the inflationary observables are simply given by
\bea \label{PR}
P_{\mathcal{R}} &=& \frac{1}{24 \pi^2 M_{\rm P}^4} \frac{V}{\epsilon}
\\ \label{ns}
n_s &=& 1 - 6 \epsilon + 2 \eta
\\
r &=& 16 \epsilon
\eea
where the slow-roll parameters are calculated as
\bea \label{ep}
\epsilon &\equiv & \frac{1}{2} \l| \frac{M_{\rm P}}{V} \frac{\partial V}{\partial I} \r|^2 
\approx \frac{\epsilon_\phi}{a^2}
\\ \label{eta}
\eta &\equiv& \frac{M_{\rm P}^2}{V} \frac{\partial^2 V}{\partial I^2} \approx \frac{\eta_\phi}{a^2}
\eea
%
\begin{figure*}[t]
\begin{center}
\includegraphics[width=0.45\textwidth]{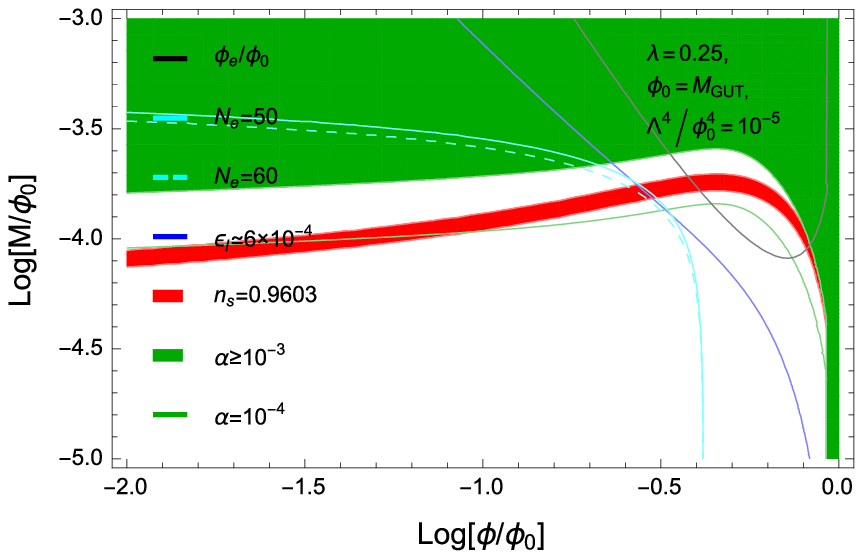}
\includegraphics[width=0.45\textwidth]{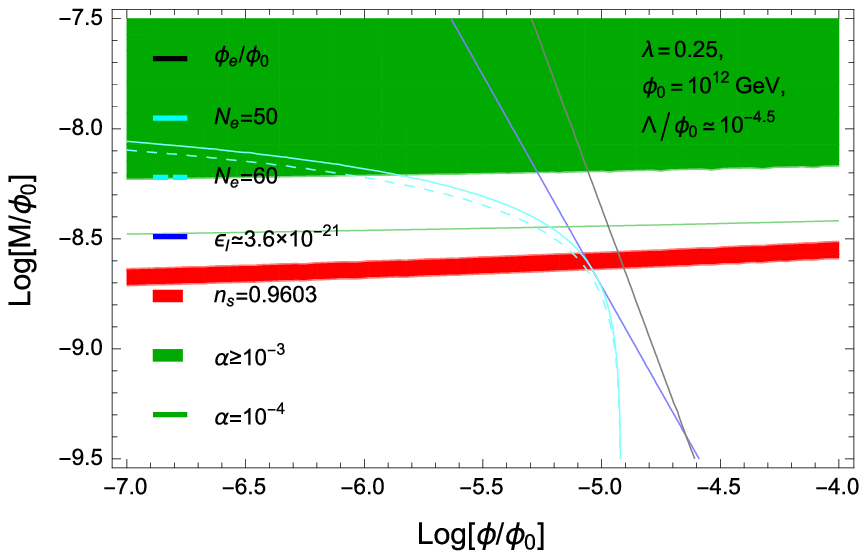}
\caption{Parameter space of inflation for $\lambda = 0.25$ with $\phi_0=M_{\rm GUT}$ (left) and $\phi_0=10^{12} \GeV$ (right).
\textit{Red}: $n_s=0.9603\pm0.007$.
\textit{Blue}: $P_{\rm R} = 2.196 \times 10^{-9}$ with $\epsilon \simeq 6 \times 10^{-4}$ (or $r\simeq0.009$) (left) and $\epsilon \simeq 3.6 \times 10^{-21}$ (right).
\textit{Green}: Shade - $n_s'>10^{-3}$, Line - $n_s'=10^{-4}$.
\textit{Cyan}: $N_e=50,60$ for dashed and solid lines, respectively.
\textit{Black}: $\phi_e/\phi_0$ as a function of $M/\phi_0$, and the $x$-axis is regarded as $\log (\phi_e/\phi_0)$ for this line.
}
\label{fig:para-space-phi0H}
\end{center}
\end{figure*}
with $\epsilon_\phi$ and $\eta_\phi$ defined as
\beq
\epsilon_\phi \equiv \frac{1}{2} \l|\frac{M_{\rm P}}{V} \frac{\partial V_\phi}{\partial \phi}\r|^2, \quad \eta_\phi \equiv \frac{M_{\rm P}^2}{V} \frac{\partial^2 V_\phi}{\partial \phi^2}
\eeq
Note again that the slow-roll parameters are suppressed by $(M/\phi)^2$ relative to the case of original CW potential.
The time derivatives of the slow-roll parameters are
\bea
\dot{\eta} &\approx& - \eta \l[ \ln \l( \frac{\phi_0}{\phi} \r) - \frac{1}{3} \r]^{-1} \l( \frac{\dot{\phi}}{\phi} \r)
\\
\dot{\epsilon} &\approx& 2 \epsilon \l[ 1 - \ln^{-1} \l(\frac{\phi_0}{\phi} \r) \r] \l( \frac{\dot{\phi}}{\phi} \r)
\eea
From \eq{dphi-dtheta-app} and the equation of motion of $\theta$, one also finds 
\beq \label{phi-dot}
\frac{\dot{\phi}}{\phi} \approx - \frac{M}{\phi} \dot{\theta} \approx \sqrt{2 \epsilon} \frac{H M M_{\rm P}}{\phi^2}
\eeq
Hence, the running of the spectral index, $\alpha \equiv d n_s/d \ln k = \dot{n}_s/\l[H \l( 1 - \epsilon \r)\r]$, is obtained as 
%
\bea \label{dns2}
\alpha 
\approx - \frac{2 \sqrt{2 \epsilon} M M_{\rm P}}{\phi^2} \l[ \frac{\eta}{\ln \l( \frac{\phi_0}{\phi} \r) - \frac{1}{3}} + 6 \epsilon - \frac{6 \epsilon}{\ln \l( \frac{\phi_0}{\phi} \r)} \r]
\eea
Note that, when $\phi$ is not very close to $\phi_0$, a negative $\alpha$ as preferred by Planck data can be otained only if $\epsilon \gtrsim \mathcal{O}(10^{-3})$ which gives $r \gtrsim \mathcal{O}(10^{-2})$ .

Since inflation ends when the CW-potential becomes too steep to hold inflaton by the modulating potential, from \eq{dVmin}, we expect the end of inflation at
\beq
\frac{\phi_e}{\phi_0} \approx \l[ \frac{1}{4 \lambda} \frac{\Lambda^4}{\phi_0^4} \l( \frac{\phi_0}{M} \r) \ln^{-1} \l( \frac{\phi_0}{\phi_e} \r) \r]^{1/3}
\eeq
which means that, for a given $M/\phi_0$, $\phi_e/\phi_0$ is controlled by $\Lambda^4/V_0$.
If $\phi \ll \phi_0$ during inflation, the number of $e$-foldings can be approximated as
\bea \label{Ne}
N_e(\phi)
&\approx& \frac{1}{M_{\rm P}} \int_\phi^{\phi_e} \frac{a^2 d\phi}{\sqrt{2 \epsilon_\phi}}
\nonumber \\ 
&\approx& - \frac{1}{\eta} \l[ 3 \ln \l( \frac{\phi_0}{\phi} \r) - 1 \r] \ln \l[ \frac{\ln (\phi/\phi_0)}{\ln (\phi_e/\phi_0)} \r]
\eea

It turns out that, as shown in the left panel of Fig.~\ref{fig:para-space-phi0H}, a CW potential with $\lambda = \mathcal{O}(0.1)$ and $\phi_0 = M_{\rm GUT}$ can give inflationary observables perfectly matching the Planck data, implying that the Higgs field breaking GUT symmetry might be responsible for the primordial inflation.
For the choice of parameters in the figure, the tensor-to-scalar ratio is $r = \mathcal{O}(0.01)$, but it can be increased if the $\phi$ relevant for the cosmological scale of interest is pushed toward $\phi_0$ (for example by taking a higher power dependence of $\phi$ in the modulating potential of \eq{VM}) with $\lambda$ increased. 
However, in the case of a large $\lambda$, the validity of perturbativity may be at risk, although the spectral running might not represent a problem. 
So, we do not consider this direction any further.

%
\begin{figure*}[t]
\centering
\includegraphics[width=0.325\textwidth]{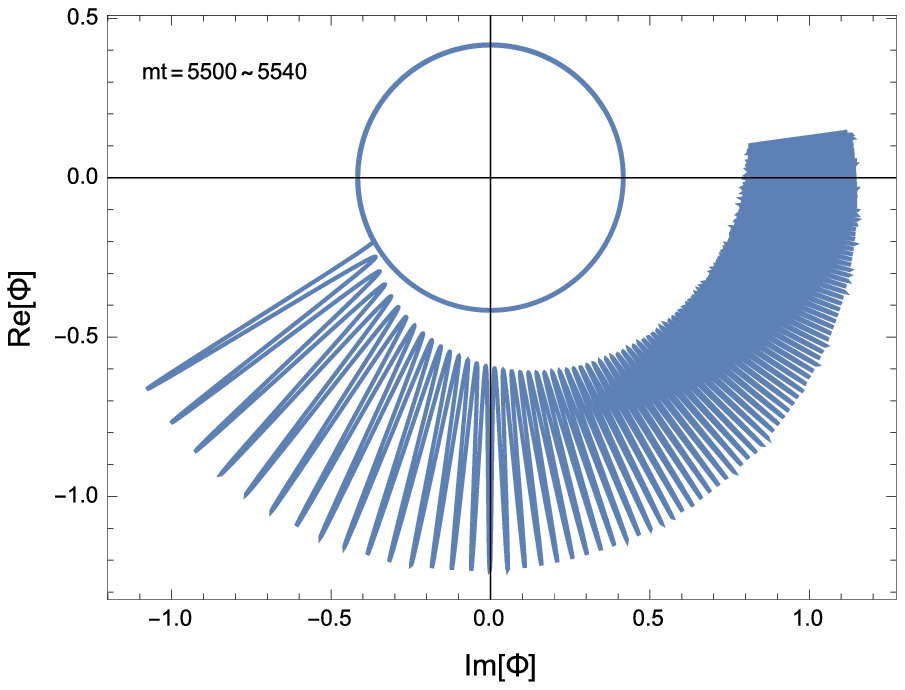}
\includegraphics[width=0.325\textwidth]{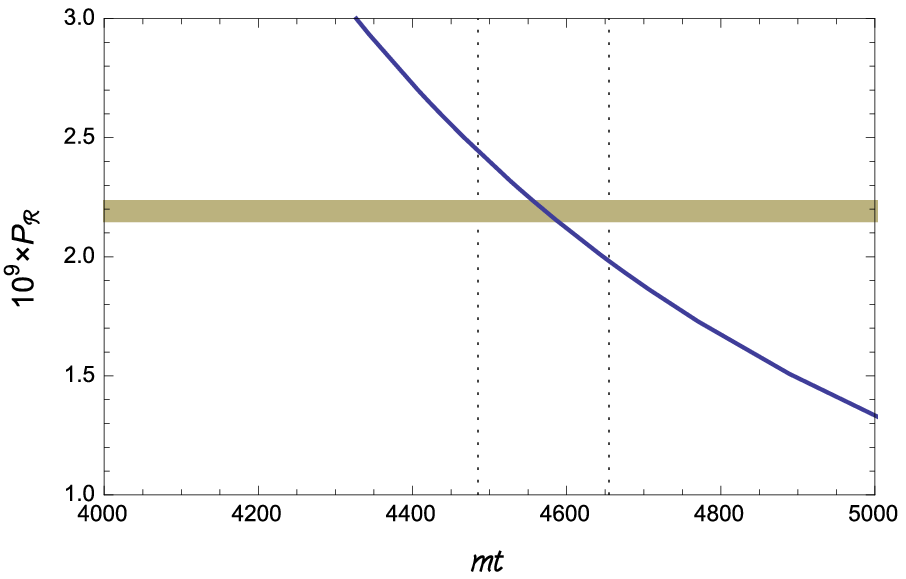}
\includegraphics[width=0.325\textwidth]{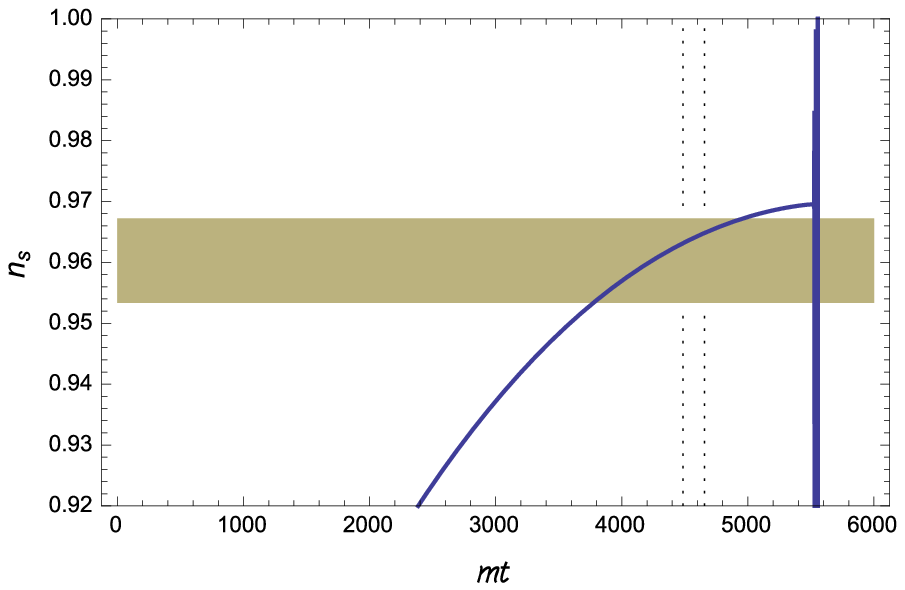}
\caption{\label{fig:observablesH} Inflaton trajectory around the end of inflation and inflationary observables for $\phi_0=M_{\rm GUT}$.
$m$ was set to be $V_0^{1/2}/(\sqrt{3} M_{\rm P} m) \simeq 6 \times 10^{-2}$.
The horizontal color band is the range allowed by Planck data at $68$ \% CL, and the vertical dotted line corresponds to $N_e=50,60$ from right to left. 
}
\end{figure*}
\begin{figure*}[t]
\centering
\includegraphics[width=0.325\textwidth]{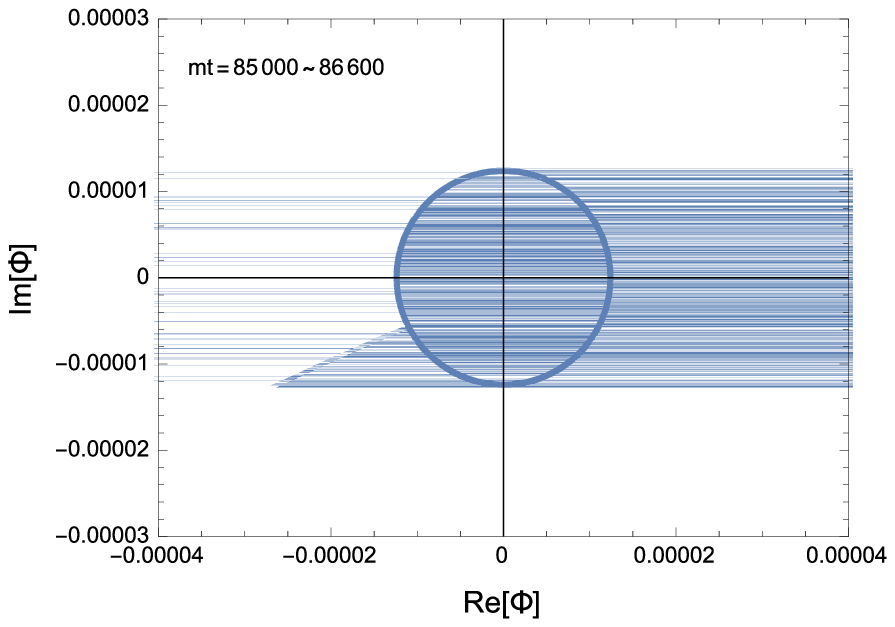}
\includegraphics[width=0.325\textwidth]{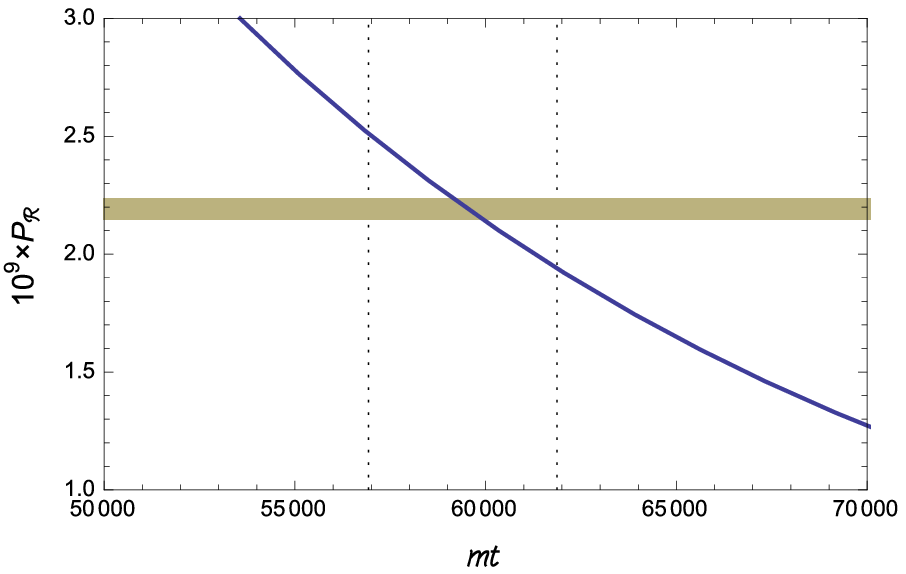}
\includegraphics[width=0.325\textwidth]{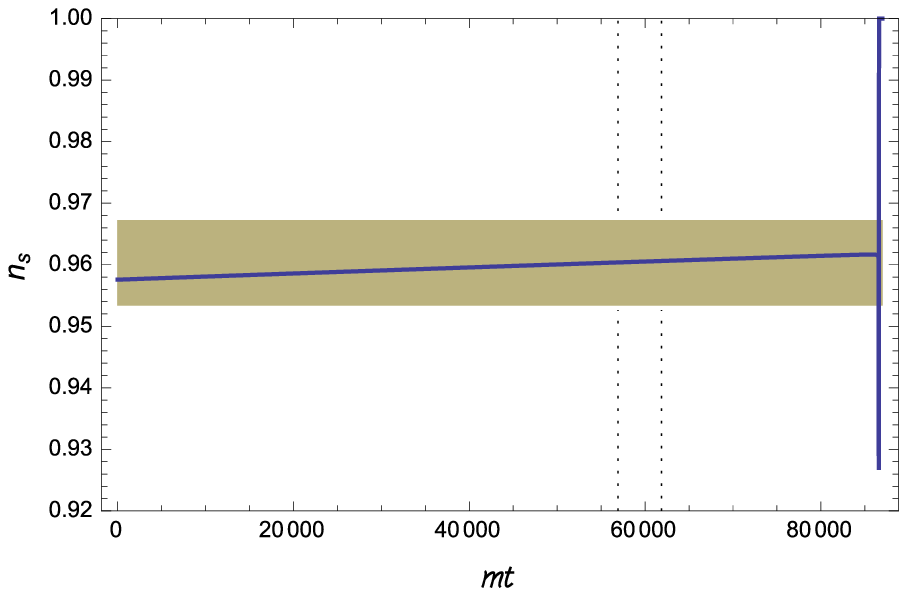}
\caption{\label{fig:observablesL} Inflaton trajectory around the end of inflation and inflationary observables for $\phi_0=10^{12} \GeV$.
$m$ was set to be $V_0^{1/2}/(\sqrt{3} M_{\rm P} m) \simeq 2 \times 10^{-3}$.
The scheme of horizontal color band and vertical dotted lines are same as Fig.~\ref{fig:observablesH}.
}
\end{figure*}
\begin{figure}[t]
\centering
\includegraphics[width=0.4\textwidth]{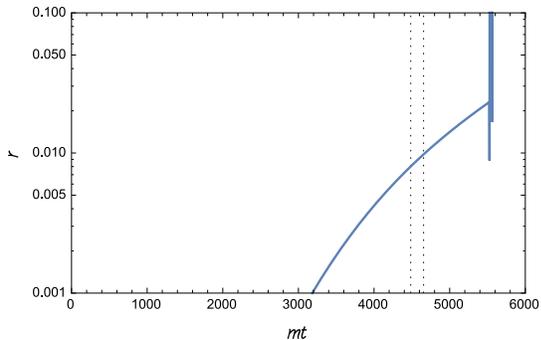}
\caption{\label{fig:rH} The tensor-to-scalar ratio for $\phi_0=M_{\rm GUT}$ with other parameters same as Fig.~\ref{fig:observablesH}.
}
\end{figure}
It is also interesting to note that a $\phi_0$ smaller than the GUT scale by several orders of magnitude can work perfectly  too, although then the tensor-to-scalar ratio turns out to be negligible.
It is easy to understand why.
When $\phi_0 \lll M_{\rm GUT}$, the observed scalar density power spectrum requires $\epsilon \lll \mathcal{O}(10^{-2})$. 
Hence, the spectral index is determined mostly by $\eta$.
Since, as can be seen from \eq{eta}, $\eta$ depends mainly on $M/\phi_0$, it is necessary to have $M/\phi_0 \sim \mathcal{O}(10^{-2}) \times \phi_0/M_{\rm P}$.
In such a case, $M \gtrsim H_I/(2 \pi)$ for $\lambda \lesssim \mathcal{O}(0.1)$.
The value of $\epsilon$ is fixed once $\lambda$ is fixed for a given $\phi_0$.
The duration of inflation or the number of $e$-foldings can be adjusted by $\Lambda^4/V_0$.
The spectral running is typically negligible since it is now of $\mathcal{O}(0.1) \times \eta^2$.
Therefore, spiral inflation on a CW potential having $\phi_0$ of around intermediate scale can work well, as shown in the right panel of Fig.~\ref{fig:para-space-phi0H} where $\phi_0=10^{12} \GeV$ was used.
Note that in the figure $\lambda$ is still of $\mathcal{O}(0.1)$, but $\Lambda/\phi_0 \simeq 10^{-4.5}$ and $M/\phi_0 \sim 10^{-8.6}$.
This hierarchy among dimensionful parameters may need a specific form of UV realization, which is out of the scope of this work. 
Also, although $\phi_e \lll \phi_0$ for an intermediate scale $\phi_0$, there is no sizable period of fast-roll inflation after spiral inflation, since the (tachyonic) mass scale is larger than the expansion rate by more than an order of magnitude.   

A remark is in order here.
In reality, there could be oscillations along $\phi$-direction, caused by the dynamics, misalignment of the initial condition, or the Hubble fluctuation, and the trajectory of the inflaton would be  slightly shifted outwards. 
However, such oscillations are expected to be damped out by the Hubble expansion.
The effect of such a deviation on slow-roll parameters, for example $\epsilon$, can be estimated as follows.
For a small enough time interval, the motion of the inflaton can be approximated roughly to an angular motion.
The associated centrifugal force should be balanced by the potential along $\phi$ direction, that is, $\partial V / \partial \phi \sim \phi \dot{\theta}^2$. 
From \eq{phi-dot}, we see that the deviation of the inflaton trajectory from the minimum of the valley gives a contribution of $\mathcal{O}(1) \times \epsilon M M_{\rm P}/\phi^2$ in the estimation of $\sqrt{2 \epsilon}$ in \eq{ep}.
It can be ignored, as can be seen from the parameter space shown in Fig.~\ref{fig:para-space-phi0H}.

\section{Numerical realization}
In this section, we demonstrate numerically the arguments discussed in the previous section.
As the testing examples of our scenario, we present two different cases with the following sets of the model parameters for $\lambda=0.25$:
\bea
{\rm I}: && \hspace{-1em} \frac{\phi_0}{M_{\rm GUT}}=1, \, \frac{\Lambda^4}{\phi_0^4} = 1.01 \times 10^{-5}, \, \frac{M}{\phi_0} = 10^{-3.8} \quad
\\
{\rm II}: && \hspace{-1em} \frac{\phi_0}{10^{12} \GeV}=1, \, \frac{\Lambda^4}{\phi_0^4} = 5.57 \times 10^{-23}, \, \frac{M}{\phi_0} = 10^{-8.6} \quad
\eea

Integrating out the equations of motions of \eqs{phi-eom}{theta-eom}, we obtained the trajectory of inflaton in each case, as shown in the left panels of Fig.~\ref{fig:observablesH} and~\ref{fig:observablesL} where power spectra and spectral indices are also depicted. 
The tensor-to-scalar ratio for $\phi_0=M_{\rm GUT}$ is shown in Fig.~\ref{fig:rH}. 
It is then clear that spiral inflation with a  CW-potential can perfectly agree with Planck data even if $\lambda = \mathcal{O}(0.1)$ with $\phi_0$ ranging from an intermediate to GUT scale.
We also found that the running of the spectral index is of $\mathcal{O}(10^{-4})$ even for $\phi_0 = M_{\rm GUT}$ which however presents a sizable tensor-to-scalar ratio.
It is experimentally indistinguishable from zero running.

\section{Conclusions}

In this paper, we showed that, when applied to a Coleman-Weinberg potential, the \textit{spiral inflation} idea allows a perfect matching to Planck data even for $\lambda = \mathcal{O}(0.1)$ and a symmetry breaking scale ranging at least from an intermediate scale to a GUT one. 
For a GUT scale symmetry-breaking, the tensor-to-scalar ratio can be of $\mathcal{O}(10^{-2})$ which may be detectable in the near future.
Even if Coleman-Weinberg potential was used here, it is easy to notice that a variety of symmetry-breaking fields appearing in low energy theories can now be used as would-be-inflaton-fields. 
For a symmetry breaking scale $\phi_0$ well below Planck scale, generically we need a mass scale of modulation ($M$) smaller than $\phi_0$ by several orders of magnitude. 
Such a hierarchy can be ameliorated if we use a higher power dependence on the symmetry breaking field in the sinusoidal modulating potential.
However, it is not clear yet how such a possibility can be realized in a UV complete theory.

\section*{Acknowledgements}
The authors acknowledge support from the MEC and FEDER (EC) Grants FPA2011-23596 and the Generalitat Valenciana under grant PROMETEOII/2013/017.
G.B. acknowledges partial support from the European Union FP7 ITN INVISIBLES (Marie Curie Actions, PITN-GA-2011-289442).


\end{document}